\documentclass[aps,pra,reprint,onecolumn,notitlepage,eqsecnum,floatfix,showkeys,nofootinbib]{revtex4-2}

\usepackage{amsmath}
\usepackage{amsbsy}
\usepackage{amsfonts}
\usepackage{bigints}
\usepackage{xcolor}
\usepackage{graphicx}
\usepackage{hyperref}
\usepackage{multirow}
\usepackage{linearb}
\usepackage{relsize}
\usepackage{amsthm}
\usepackage{slashed}
\usepackage{footmisc}
\usepackage{calrsfs}
\DeclareMathAlphabet\mathbfcal{OMS}{cmsy}{b}{n}

\newcommand{\bq}{\begin{eqnarray}}
\newcommand{\eq}{\end{eqnarray}}
\newcommand{\bqn}{\begin{eqnarray*}}
\newcommand{\eqn}{\end{eqnarray*}}
\newcommand{\bqs}{\begin{subequations}}
\newcommand{\eqs}{\end{subequations}}
\newcommand{\bw}{\begin{widetext}}
\newcommand{\ew}{\end{widetext}}
\newcommand{\xx}{{\boldsymbol x}}

\newcommand{\aaa}{{\boldsymbol a}}

\newcommand{\nn}{{\boldsymbol n}}
\newcommand{\rr}{{\boldsymbol r}}
\newcommand{\zero}{{\boldsymbol 0}}
\newcommand{\uu}{{\boldsymbol u}}
\newcommand{\vv}{{\boldsymbol v}}
\newcommand{\zz}{{\boldsymbol z}}

\newcommand{\pp}{{\boldsymbol p}}

\newcommand{\nnabla}{{\boldsymbol\nabla}}

\newcommand{\calm}{{\cal M}}
\newcommand{\calo}{{\cal O}}

\newcommand{\calp}{{\cal P}}

\newcommand{\calh}{{\cal H}}

\newcommand{\calv}{{\cal V}}
\newcommand{\calk}{{\cal K}}
\newcommand{\caln}{{\cal N}}

\newcommand{\vect}[1]{\vec{\boldsymbol #1}}
\newcommand{\form}[1]{\widetilde{\boldsymbol #1}}
\newcommand{\gig}{{\boldsymbol g}\!\!{\boldsymbol g}}

\newcommand{\red}[1]{{#1}}

\begin{document}
\title{Many Body in General Relativity: A thermal equivalence principle}

\author{Riccardo Fantoni}
\email{riccardo.fantoni@scuola.istruzione.it}
\affiliation{Universit\`a di Trieste, Dipartimento di Fisica, strada
  Costiera 11, 34151 Grignano (Trieste), Italy}

\date{\today}

\begin{abstract}
We review the physics of many bodies in the context of general relativity.
Starting from the stress energy tensor for one body, for a swarm of bodies,
for a perfect fluid, we review relativistic hydrodynamics, kinetic theory, 
and statistical physics of $N$ identical bodies. We conclude our excursion 
with a {\sl thermal equivalence principle} in physics.
\end{abstract}

\keywords{Particle; Swarm; Perfect Fluid; Hydrodynamics; Kinetic Theory; Statistical Mechanics; General Relativity; Equivalence Principle}

\maketitle
\tableofcontents
\section{Introduction}
\label{sec:intro}

An interesting problem in physics is to study the properties of a (quantum) 
{\sl many body} system at low (non-zero) temperature on a {\sl curved surface}. 
For example colloidal particles may be adsorbed or confined on a substrate 
with nonzero curvature, be it the wall of a porous material, or a membrane, a
vesicle, a micelle for example made of ampiphilic surfactant molecules
such as lipids, or a biological membrane, or the surface of a large
solid particle, or an interface in an oil-water emulsion \cite{Fantoni12c}.
For a fluid of $^4$He atoms it would be interesting to study the superfluidity. 
For a fluid of electrons it would be interesting to study the 
\red{superconductivity}. 

One important point to discuss is whether the space in which the particles
live is exactly two dimensional, as it happens in the satirical novella of 
Edwin Abbott Abbott \cite{Abbott}, or if it can be treated as {\sl quasi}
two dimensional. There is a profound difference between the two scenarios 
to the point that the form of the interaction between the particles also
changes. For example for colloidal particles one may choose the polarizable 
hard sphere pair interaction or for the fluid of helium atoms one may use 
the Lennard Jones pair potential, but the distance between the two 
interacting particles may be chosen either as the geodesic distance between 
them or the Euclidean distance in the three dimensional space where the 
surface is embedded. 
For the electron gas the Coulomb pair potential as a solution to the 
Poisson equation has different forms in two or three dimensions and in 
general depends on the metric of the curved surface \red{\cite{Fantoni19a}}.

\red{More generally it would be desirable to advance our knowledge for 
treating a quantum many body system in a Riemannian space. For example 
this would make possible refinements of the equations of state of stars, 
like for example a White Dwarf, in our universe ruled by general relativity
laws and described in terms of a curved spacetime.}

These properties can be studied exactly with the path integral 
(Monte Carlo) method and these studies certainly enrich the knowledge on 
many bodies in (quantum) general relativity
\cite{Fantoni24f, Fantoni25a, Fantoni25g, FantoniCQ}. 
Not even the two body problem can be treated analytically in general 
relativity \cite{Fantoni25f}. The problem of gravitating many bodies
should be separated by the problem of many bodies with non-gravitational
interactions in general relativity. In fact mass curves
spacetime through the Einstein field equations and gravitating bodies
will behave as free particles on that curved spacetime, whereas 
non-gravitational interactions produce particles accelerations
on the spacetime. So being able to treat many (quantum) bodies on
a curved surface would be an important step forward for the much more
complicated problem of gravitating many (quantum) bodies in
general relativity.

We find it of fundamental importance issuing a bridge between the two 
scientific communities of the exact simulations of a many body (quantum) 
system and of general relativity. We foresee an important progress in the 
physics of (quantum) gravitating many body systems beyond the simple ideal 
gases or hydrodynamic systems that are usually treated 
\cite{Shapiro-Teukolsky, Gravitation}. We therefore review the physics of many 
bodies in the context of general relativity.
Starting from the stress energy tensor for one body, for a swarm of bodies,
for a perfect fluid, we review relativistic hydrodynamics, kinetic theory, 
and statistical physics of $N$ {\sl identical} bodies. We conclude our 
excursion with a {\sl thermal equivalence principle} in physics \red{that 
proves the consistency between statistical physics and general relativity}.

In this work we consider {\sl spacetime} as a smooth manifold $\calm$ of 
dimension $d$ and metric tensor $\gig$
with covariant components $g_{\alpha\beta}$. We will denote with an arrow
over a bold face letter the corresponding 4-vector and with just the bold face
symbol the corresponding 3-dimensional vector. Greek indexes run over the $d$ 
spacetime dimensions. Roman indexes run only over the $d-1$ space dimensions. 
We use Einstein summation convention of tacitly assuming a sum over repeated 
indexes. We will assume the speed of light $c=1$ throughout.

\red{The paper is organized as follows. In the first three sections we review 
well known concepts in general relativity. In Section \ref{sec:setf} we
give the definition of the stress energy tensor for a free particle, a swarm 
of free particles, and a perfect fluid; in Section \ref{sec:hydro} we describe 
the laws of Newtonian and Relativistic hydrodynamics; in Section 
\ref{sec:kinetict} we describe the kinetic theory approach in general 
relativity. In the last Section \ref{sec:statm} we propose a statistical 
physics approach in general relativity demonstrating how the two theories
of statistical (quantum) theory and of general relativity are consistent
among themselves. This last section constitute the main novel result of 
the work presenting a new thermal equivalence principle. 

\section{The stress energy tensor for free particles}
\label{sec:setf}
}
Here we review the basic definition of the stress energy tensor of general 
relativity for a free particle, a swarm of free particles, and a
perfect fluid.

\subsubsection{One particle}
\label{sec:onep}

For {\sl one body} of mass $m$ we have a self gravitating system with a 
stress energy tensor given by
\bq
T^{\alpha\beta}(\vect{\xx})=m\int u^\alpha u^\beta
\delta^{(4)}(\vect{\xx}-\vect{\zz}(\tau))\,d\tau,
\eq
where $\tau$ is the body proper time, 
$d\vect{\zz}/d\tau=\vect{\uu}=(\gamma,\gamma\vv)$ with 
$u^0=dt/d\tau=\gamma=(1-v^2)^{-1/2}$ and
\bq
T^{\alpha\beta}(\vect{\xx})=m\frac{u^\alpha u^\beta}{u^0}
\delta^{(3)}(\xx-\zz(t)),
\eq
where the body is at $\zz(t)$ with velocity $\vv(t)$ at time $t$.

\subsubsection{Swarm of particles}
\label{sec:swarm}

For a {\sl swarm} of $N$ bodies all of the same mass $m$ and $\vv$
\bq \nonumber
T^{\alpha\beta}(\vect{\xx})&=&m u^\alpha u^\beta\sum_{i=1}^N
\int\delta^{(4)}(\vect{\xx}-\vect{\zz}_i(\tau_i))\,d\tau_i\\ \nonumber
&=&m\frac{u^\alpha u^\beta}{u^0}\sum_{i=1}^N\delta^{(3)}(\xx-\zz_i(t))\\
&=&m u^\alpha u^\beta n,
\eq
where
\bq
n=\frac{1}{u^0}\sum_{i=1}^N\delta^{(3)}(\xx-\zz_i(t)),
\eq
is the proper number density of bodies measured in a {\sl comoving frame}
where $\vect{u}=(1,\zero)$.

\subsubsection{Perfect fluid}
\label{sec:pfluid}

For a {\sl perfect fluid} of proper number density $n$ of non interacting bodies 
all of the same mass $m$ and $v=|\vv|$ but isotropic velocity profile $\vv=v\nn$
\bq
T^{\alpha\beta}=\chi\langle u^\alpha u^\beta\rangle_\nn,
\eq
so that $T^{\alpha\beta}=0$ for $\alpha\neq\beta$ and $T^{00}=\chi\gamma^2$.
Since $T^{00}=\rho=n(\gamma m)$ is the energy density of the fluid we require
$\chi=mn/\gamma$. Then
\bq \nonumber
T^{ij}&=&\chi\gamma^2 v^2\langle n^in^j\rangle_\nn\\ \nonumber
&=&\chi\gamma^2 v^2\frac{1}{3}\delta^{ij}\\ \nonumber
&=&n(\gamma m)v^2\frac{1}{3}\delta^{ij}\\
&=&p\delta^{ij},
\eq
where $\delta$ is a Kronecker delta and in the second equality we used 
isotropy of $\nn$ and 
\bq
\left\{\begin{array}{l}
\rho=n(m\gamma)\\
p=\frac{1}{3}\rho v^2
\end{array}\right.
\eq
are respectively the mass density and pressure in the {\sl isotropic frame} 
of the fluid. Summarizing
\bq \label{eq:setpf}
T^{\alpha\beta}=(\rho+p)u^\alpha u^\beta + p\eta^{\alpha\beta},
\eq
where $||\eta^{\alpha\beta}||={\rm diag}\{-1,1,1,1\}$ is the metric in 
Minkowski spacetime.
For photons $v=1$ and $p=\rho/3$. For $v\ll 1$, $\rho=nm(1+v^2/2+\ldots)$,
and $p\approx nmv^2/3=(2/3)(\rho-nm)=(2/3)\epsilon$, where 
$\epsilon=(3/2)k_B T$ is the internal energy of a monatomic ideal gas in 
thermal equilibrium at a temperature $T$, $k_B$ is Boltzmann constant, and
$p=nk_B T$ is the ideal gas equation of state.
 
\section{Hydrodynamics}
\label{sec:hydro}

{\sl Hydrodynamics} concerns itself with the study of the motion of fluids
(liquids and gases). Since the phenomena considered in fluid dynamics are 
macroscopic, a fluid is regarded as a continuous medium. Therefore when we
speak of the ``point'' of a fluid (or of an infinitesimal volume of it) we 
mean not a single molecule of the fluid but a volume element still containing 
very many molecules but yet small compared with the volume of the whole fluid.

\subsection{Newtonian}
\label{sec:hydron}

A mathematical description of the state of a moving fluid consists in 
specifying the fluid velocity $\vv=\vv(t,\xx)$ and any two thermodynamic 
functions pertaining to the fluid, for instance the pressure 
$p=p(t,\xx)$ and the density $\rho=\rho(t,\xx)$, from which one can determine 
all other thermodynamic quantities. These 5 quantities are functions of the 
coordinates $\xx=(x,y,z)$ and of time $t$. Once again we stress that a point 
$\rr$ in space at a given time $t$ refers to a fixed point and not to specific 
particles of the fluid. From Chapter I of Ref. \cite{LandauFM} we find
\bq \label{eq:nce}
&&\frac{\partial\rho}{\partial t}+\nnabla(\rho\vv)=0,\\ \label{eq:nee}
&&\frac{\partial\vv}{\partial t}+(\vv\cdot\nnabla)\vv=-\frac{1}{\rho}\nnabla p,\\ \label{eq:naf}
&&\frac{\partial s}{\partial t}+(\vv\cdot\nnabla)s = 0,
\eq
where the first equation is the {\sl continuity equation}, the second is the
{\sl Euler equation}, and the third one is the {\sl equation for the adiabatic 
flow} in which $s=s(t,\xx)$ is the entropy per particle.

From the first law of thermodynamics follows
\bq \label{eq:nflt}
&&d\epsilon=T\,ds-p\,d(m/\rho),\\ \label{eq:nie}
&&\epsilon=\epsilon(\rho,s),\\ \label{eq:npressure}
&&p=\frac{\rho^2}{m}\left.\frac{\partial\epsilon}{\partial\rho}\right|_s,\\ \label{eq:nT}
&&T=\left.\frac{\partial\epsilon}{\partial s}\right|_\rho,
\eq
where $\epsilon$ is the internal energy per particle. Eqs. (\ref{eq:npressure}) 
and (\ref{eq:nT}) can be considered as algebraic relations for the right hand 
side of Eqs. (\ref{eq:nee}) and (\ref{eq:naf}) respectively.

For an ideal gas $\epsilon=\epsilon(T)$ and for a monatomic gas
\bq
s=k_B\ln(T^{3/2}m/\rho)+\mbox{constant}.
\eq

\subsection{Relativistic}
\label{sec:hydror}

We will work in a Local Lorentz Frame (LLF). Recalling that the stress energy 
tensor is divergenceless, from the stress energy tensor of a perfect fluid 
(\ref{eq:setpf}) we find
\bq \nonumber
0=T^{\alpha\beta}{}_{,\beta}&=&(\rho+p)_{,\beta} u^\alpha u^\beta+
(\rho+p)u^\alpha{}_{,\beta}u^\beta +
(\rho+p)u^\alpha u^\beta{}_{,\beta} + p_{,\beta}\eta^{\alpha\beta}\\
&=&\frac{d(\rho+p)}{d\tau}u^\alpha+(\rho+p)a^\alpha+(\rho+p)
u^\alpha u^\beta{}_{,\beta}+p_,{}^\alpha,
\eq
where the comma stands for a partial derivative. 
Multiplying by $\vect{\uu}$ and recalling that $\vect{\uu}\cdot\vect{\aaa}=0$
we find
\bq \label{eq:rce}
\frac{d\rho}{d\tau}=-(\rho+p)\vect{\nnabla}\vect{\uu},
\eq
which is the relativistic continuity expression which extends Eq. (\ref{eq:nce}).

To find the extension of the Euler equation we introduce the projector
tensor
\bq \nonumber
P^{\alpha\beta}&=&\eta^{\alpha\beta}+u^\alpha u^\beta
~~~\mbox{for $\vect{\uu}$ timelike}~~~\vect{\uu}\cdot\vect{\uu}=-1\\ \nonumber
P^{\alpha\beta}&=&\eta^{\alpha\beta}-n^\alpha n^\beta
~~~\mbox{for $\vect{\nn}$ spacelike}~~~\vect{\nn}\cdot\vect{\nn}=+1
\eq
Then
\bq
0=P_{\alpha\gamma}T^{\alpha\beta}{}_{,\beta}=
(\rho+p)a_\gamma+P_{\alpha\gamma}p_,{}^\alpha,
\eq
or
\bq \label{eq:ree}
(\rho+p)\vect{\aaa}=-\vect{\nnabla}p-\vect{\uu}\frac{dp}{d\tau},
\eq
which is the relativistic Euler equation which extends Eq. (\ref{eq:nee}). 

It is easy to see that in the non relativistic limit 
$\vect{\uu}=(\gamma,\gamma\vv)\approx (1,\vv)$ with $v\ll 1$ and 
$p\ll\rho$, Eq. (\ref{eq:rce}) reduces to Eq. (\ref{eq:nce}) and 
Eq. (\ref{eq:ree}) reduces to Eq. (\ref{eq:nee}) \cite{HTWW}
\footnote{To determine the stability of a star it is often sufficient to 
replace Eq. (\ref{eq:ree}) with Eq. (\ref{eq:nee}) as reported in 
\S 6.9 of Ref. \cite{Shapiro-Teukolsky}.}.

Let us now discuss the continuity Eq. (\ref{eq:rce}). First of all we observe 
that the mass density is not conserved $d\rho/d\tau\neq 0$. But the baryon, 
lepton, charge, $\dots$ numbers are conserved. For example if we
call $n=N/V$ the baryon number density in the rest frame of the fluid with
$N$ baryons in a volume $V$, $N$ is certainly constant but $V$ will change,
so that 
\bq
0=\frac{dN}{d\tau}=\frac{d(nV)}{d\tau},
\eq
but $(dV/d\tau)/V=\vect{\nnabla}\vect{\uu}$ (see Ex. 22.1 in Ref. 
\cite{Gravitation}). So 
\bq \nonumber
0&=&\frac{1}{V}\frac{d(nV)}{d\tau}\\ \nonumber
&=&\frac{dn}{d\tau}+n\vect{\nnabla}\vect{\uu}\\ \nonumber
&=&\vect{\uu}\cdot\vect{\nnabla}n+n\vect{\nnabla}\vect{\uu}\\ \label{eq:rce1}
&=&\vect{\nnabla}(n\vect{\uu}),
\eq
where we may define the divergenceless current density
\bq \label{eq:cd}
\vect{{\boldsymbol J}}=n\vect{\uu}.
\eq

Let us now discuss the thermodynamics. The second law tells that $ds/d\tau\geq 0$
where $s$ is the entropy per baryon. The first law becomes
\bq
d(\rho/n)=-p\,d(1/n)+T\,ds,
\eq
or
\bq \label{eq:rflt}
d\rho=\frac{\rho+p}{n}\,dn+nT\,ds,
\eq
which is the relativistic extension of Eq. (\ref{eq:nflt}). In this equation
the differential $d$ can be substituted either with an exterior derivative 
$\form{d}$, with a gradient $\vect{\nnabla}$, or with a directional derivative
$\vect{\nnabla}_{\vect{\uu}}=u^\alpha\partial/\partial x^\alpha=d/d\tau$. 
Given an equation of state $\rho=\rho(n,s)$ we will have
\bq \label{eq:rpressure}
&&p=n\left.\frac{\partial\rho}{\partial n}\right|_s-\rho,\\ \label{eq:rT}
&&T=\frac{1}{n}\left.\frac{\partial\rho}{\partial s}\right|_n,
\eq
which are the relativistic extensions of Eqs. (\ref{eq:npressure}) and 
(\ref{eq:nT}).

It is easy to show that a perfect fluid flow is adiabatic. From the 
relativistic continuity Eqs. (\ref{eq:rce}) and (\ref{eq:rce1}) follows
\bq
\frac{d\rho}{d\tau}=\frac{\rho+p}{n}\frac{dn}{d\tau}.
\eq
Then from the relativistic first thermodynamic Eq. (\ref{eq:rflt}) follows
\bq
\frac{ds}{d\tau}=0.
\eq

\subsubsection{Shock wave}

Consider a homogeneous, static, perfect fluid. A sound wave in the fluid is an 
adiabatic perturbation. The speed of sound is
\bq
v_s^2=\left.\frac{\partial p}{\partial\rho}\right|_s
\eq

Expand
\bq \nonumber
\rho&=&\rho_0+\rho_1,\\ \nonumber
p&=&p_0+p_1,\\ \nonumber
n&=&n_0+n_1,
\eq
where $\rho_0, p_0, n_0$ are constant in space (uniform fluid) and in time
(static fluid) and $\rho_1, p_1, n_1$ are small perturbations. 
Taking $\vect{\uu}=(1,\vv_1)$ with $v_1\ll 1$ we find from the continuity
Eq. (\ref{eq:rce})
\bq \label{eq:s1}
\frac{\partial\rho_1}{\partial t}=-(\rho_0+p_0)\nnabla\vv_1,
\eq
and from the spatial part of Euler Eq. (\ref{eq:ree}) 
\bq \label{eq:s2}
(\rho_0+p_0)\frac{\partial\vv_1}{\partial t}=-\nnabla p_1,
\eq
where we neglect the last term 
$\vect{\uu}\,dp/d\tau=\vect{\uu}\,u^\alpha\partial p/\partial x^\alpha$ because
an infinitesimal of second order and $\partial p_0/\partial t=0$. Therefore 
putting together Eqs. (\ref{eq:s1}) and (\ref{eq:s2}) we find
\bq
\frac{\partial^2\rho_1}{dt^2}=-(\rho_0+p_0)\nnabla\frac{\partial\vv_1}{\partial t}
=\nnabla^2 p_1.
\eq
In a perfect fluid $p=p(\rho,T)$ so that $p(\rho_0+\rho_1,T)=p(\rho_0,T)+
\partial p(\rho_0,T)/\partial\rho|_s\rho_1=p_0+p_1$ with $p_1=v_c^2\rho_1$ and 
we finally find
\bq
\frac{\partial^2\rho_1}{dt^2}=v_s^2\nnabla^2\rho_1,
\eq 
which is the shock wave equation.

\subsubsection{Bernoulli equation}
Consider a steady, adiabatic flow of a perfect fluid. Since in a steady
state $\partial p/\partial t=0$ from the relativistic Euler Eq. (\ref{eq:ree})
follows
\bq \label{eq:ree1}
(\rho+p)\frac{du^0}{d\tau}=-u^0\frac{dp}{d\tau},
\eq
So
\bq \nonumber
\frac{d}{d\tau}\left(u^0\frac{\rho+p}{n}\right)&=&
\frac{du^0}{d\tau}\frac{\rho+p}{n}+u^0\frac{d}{d\tau}
\left(\frac{\rho+p}{n}\right)\\ \nonumber
&=&-\frac{u^0}{n}\frac{dp}{d\tau}+u^0\frac{d}{d\tau}
\left(\frac{\rho+p}{n}\right)\\
&=&\frac{u^0}{n}\left[\frac{d\rho}{d\tau}-
\frac{\rho+p}{n}\frac{dn}{d\tau}\right]=0,
\eq
where in the second equality we used Eq. (\ref{eq:ree1}) and in the last 
equality we used the relativistic first law of thermodynamics
Eq. (\ref{eq:rflt}). We then conclude that 
\bq
u^0\frac{\rho+p}{n}~~~\mbox{is constant along the fluid flow lines.}
\eq

In the Newtonian limit 
$\vect{\uu}=(\gamma,\gamma\vv)\approx (1,\vv)$ with $v\ll 1$, 
$u^0=\gamma=(1-v^2)^{-1/2}\approx 1+v^2/2$, and $P\ll\rho$. We can take
$\rho=\rho_0(1+\pi)$ and 
$(\rho+p)/n\approx\rho_0(1+\pi+p/\rho_0)/n$, so that
\bq
\frac{1}{2}v^2+\pi+p/\rho_0~~~\mbox{is constant along the fluid flow lines},
\eq
where the sum of the last two terms is the {\sl enthalpy}.

\section{Kinetic theory approach}
\label{sec:kinetict}

The kinetic theory approach is based on a {\sl one body} distribution function
\cite{Hansen,Gravitation}.

\subsubsection{Distribution function}

We will construct a Lorentz invariant phase space distribution function as a 
number density of particles in phase space
\bq
f=\frac{d\caln}{d\xx\,d\pp},~~~~~~\int f\,d\xx\,d\pp = N,
\eq
for a fluid of $N$ bodies, where $d\xx=dx^1dx^2dx^3$ and 
$d\pp=dp^1dp^2dp^3$. So that $f/N$ can be considered as a probability 
distribution function. We will now prove that $f$ as defined above is a 
Lorentz invariant distribution. We start defining a proper 3-volume. The 
4-volume $d^4\Omega=dx^0\,dx^1\,dx^2\,dx^3$ is invariant under a Lorentz 
transformation. Dividing by $d\tau$ we find another Lorentz invariant
\bq
dV=u^0\,dx^1\,dx^2\,dx^3.
\eq
Then we want to define a 3-volume element in momentum space. The 
4-volume $d^4p=dp^0\,dp^1\,dp^2\,dp^3$ is invariant. Since 
$p^0=\sqrt{\pp^2+m^2}$ we will define
\bq \nonumber
d\Pi&=&\int d^4p\,\delta\left(\sqrt{-\vect{\pp}\cdot\vect{\pp}}-m\right)\\
\nonumber
&=&\frac{\sqrt{(p^0)^2-\pp^2}}{p^0}dp^1\,dp^2\,dp^3\\
&=&\frac{m}{p^0}\,dp^1\,dp^2\,dp^3.
\eq
And
\bq
dV\,d\Pi=dx^1\,dx^2\,dx^3\,\,dp^1\,dp^2\,dp^3,
\eq
is Lorentz invariant.

We will now prove conservation of volume in phase space in curved spacetime 
(see Liouville theorem in BOX 22.6 of Ref. \cite{Gravitation}). Consider a 
very small bundle of identical particles that move through curved spacetime 
on a neghboring geodesics. We want to prove that $d(dV\,d\Pi)/d\lambda=0$
where $\lambda$ is an affine parameter along the central geodesic of the bundle.
Given any function of phase space $g(\vect{\xx},\vect{\pp})$, if $m\neq 0$
\footnote{If $m=0$ see BOX 22.6 of Ref. \cite{Gravitation}.} take 
$\tau=a\lambda+b$ for arbitrary $a$ and $b$. Then
\bq
\frac{dg(\vect{\xx},\vect{\pp})}{d\lambda}=
\frac{\partial g}{\partial x^\alpha}\frac{dx^\alpha}{d\lambda}+
\frac{\partial g}{\partial p^\alpha}\frac{dp^\alpha}{d\lambda},
\eq
on a geodesic $dp^\alpha/d\tau=0$ so
\bq
\frac{dg(\vect{\xx},\vect{\pp})}{d\lambda}=
\frac{\partial g}{\partial x^\alpha}p^\alpha\frac{a}{m},
\eq
and for $g=dV\,d\Pi$ and $p^\alpha=mdx^\alpha/d\tau$
\bq
\frac{dg}{d\lambda}=a\frac{dg}{d\lambda},
\eq
for any $a$. so $dg/d\lambda=0$. Since $d\caln$ and $dV\,d\Pi$ are unchanged
then also $f=d\caln/dV\,d\Pi$ is unchanged
\bq \label{eq:Liouville}
\frac{df}{d\lambda}=0.
\eq

This equation is at the heart of the collisionless Boltzmann equation and the 
Vlasov equation. All these approximate theories are valid at sufficiently low 
density. Whereas  for the full Boltzmann equation \cite{Resibois1977,Hansen} one 
has
\bq
\frac{df}{d\lambda}=\left(\frac{\partial f}{\partial\lambda}\right)_{\rm collisions}.
\eq 
This is the most famous of all Kinetic equations and was obtained by 
Boltzmann more than a century ago. 

The phase space probability density of a system in thermodynamic equilibrium
at an inverse temperature $\beta=1/k_BT$ with $k_B$ Boltzmann constant, is not an
explicit function of proper time.  We shall use the symbol $f_0$ to denote
the equilibrium probability density.

\subsubsection{Ideal gas}

For an {\sl ideal gas}, i.e. a fluid of non interacting many identical bodies, 
from \S 37 \& 52 \& 53 of Ref. \cite{LandauSP} we know that, on a comoving 
frame with $\vect{\uu}=(1,\zero)$, we can write 
\bq \label{eq:obpdf}
f_0(\vect{\xx},\vect{\pp})=\frac{d\caln}{dV\,d\Pi}=
\frac{g}{h^3}\frac{1}{e^{-\beta(\vect{\pp}\cdot\vect{\uu}+\mu)}-\varepsilon},
\eq
where $h$ is Planck constant, $\mu$ is the chemical potential, $g=2J+1$ is the 
spin $J$ degeneracy (2 polarizations for photons), and
\bq
\varepsilon=\left\{\begin{array}{ll}
+1 & \mbox{Bose-Einstein statistics}\\
0 & \mbox{Maxwell-Boltzmann statistics}\\
-1 & \mbox{Fermi-Dirac statistics}
\end{array}\right.
\eq
Where $\varepsilon=0$ takes care of the statistics for classical 
{\sl distinguishable} bodies at sufficiently high temperature. At sufficiently low 
temperature a different statistics must be devised, in which the mean occupation 
number of the various quantum states of bodies are not assumed small. The 
statistics, however, differs according to the type of many body wave function 
by which the gas is described. These functions must be either symmetrical or
antisymmetrical with respect to interchange of any pair of particles 
(see \S 61 in Ref. \cite{LandauQM}). The former case occurring for bodies 
with integral spin, {\sl bosons} $\varepsilon=+1$, and the latter case for those of 
half-integral spin, {\sl fermions} $\varepsilon=-1$. 

\subsubsection{Moments of the distribution function of the ideal gas}

Next we can take moments of $f_0$ with respect to $\vect{\pp}$
\bq
\int f_0p^\mu\,d\Pi&=&J^\mu,\\
\int f_0p^\mu p^\nu\,d\Pi&=&T^{\mu\nu},
\eq
where here $d\Pi=d\pp/p^0$ with $p^0=\sqrt{\pp^2+m^2}$. For $\vect{\uu}=(1,\zero)$,
from Eqs. (\ref{eq:cd}) and (\ref{eq:setpf}), we must have
\bq
J^\mu&=&nu^\mu,\\
T^{\mu\nu}&=&(\rho+p)u^\mu u^\nu+p\eta^{\mu\nu}.
\eq
So
\bq \nonumber
n&=&-J^\mu u_\mu=-\int f_0p^\mu u_\mu\,d\Pi=\int f_0\,d\pp\\ \label{eq:ne1}
&=&\frac{g}{h^3}\int_0^\infty\frac{4\pi p^2\,dp}
{e^{\beta[\sqrt{\pp^2+m^2}-\mu]}-\varepsilon}.
\eq
Introduce the following change of variables
\bq
\left\{\begin{array}{l}
p=m\sinh\chi\\
\bar{\beta}=m\beta
\end{array}\right.
\eq
so that from Eq. (\ref{eq:ne1}) we find
\bq \label{eq:ne2}
n=\frac{4\pi g m^3}{h^3}\int_0^\infty\frac{\sinh^2\chi\,\cosh\chi\,d\chi}
{e^{[\bar{\beta}\cosh\chi-\beta\mu]}-\varepsilon}.
\eq
For the pressure
\bq \nonumber
p&=&\frac{1}{3}(u_\mu u_\nu+\eta_{\mu\nu})T^{\mu\nu}\\ \nonumber
&=&\frac{1}{3}\int f_0\pp^2\,d\Pi\\ \nonumber
&=&\frac{1}{3}\int f_0\pp^2\,\frac{d\pp}{p^0}\\ 
&=&\frac{4\pi gm^4}{3h^3}\int_0^\infty\frac{\sinh^4\chi\,d\chi}
{e^{[\bar{\beta}\cosh\chi-\beta\mu]}-\varepsilon}.
\eq
Also
\bq \nonumber
\rho-3p&=&-T^\alpha{}_\alpha=m^2\int f_0\,\frac{d\pp}{p^0}\\
&=&\frac{4\pi gm^4}{h^3}\int_0^\infty\frac{\sinh^2\chi\,d\chi}
{e^{[\bar{\beta}\cosh\chi-\beta\mu]}-\varepsilon}.
\eq

\subsubsection{Maxwell-Boltzmann statistics ($\varepsilon=0$) \cite{Sharp2015}}
From Ref. \cite{Abramowitz} we learn that
\bq
K_n(\bar{\beta})&=&\frac{\bar{\beta}^n}{(2n-1)!!}\int_0^\infty 
d\chi\,\sinh^{2n}\chi\, e^{-\bar{\beta}\cosh\chi}\\
&=&\frac{\bar{\beta}^{n-1}}{(2n-3)!!}\int_0^\infty 
d\chi\,\sinh^{2n-2}\chi\cosh\chi\, e^{-\bar{\beta}\cosh\chi},
\eq
where $K_n$ is a modified Bessel function of the second kind and in the second 
equality we performed an integration by parts. The asymptotic behaviors of 
the modified Bessel function are as follows
\bq \label{eq:asy1}
K_n(\bar{\beta})&=&\sqrt{\frac{\pi}{2\bar{\beta}}}e^{-\bar{\beta}}
\left[1+\frac{4n^2-1}{8\bar{\beta}}+O(\bar{\beta}^{-2})\right]
~~~\bar{\beta}\gg 1,\\ \label{eq:asy2}
K_n(\bar{\beta})&=&\frac{(n-1)!}{\bar{\beta}^n}
\left[2^{n-1}-\frac{2^{n-3}\bar{\beta}^2}{n-1}+O(\bar{\beta}^3)\right]
~~~\bar{\beta}\ll 1.
\eq

We then find
\bq \label{eq:nmb}
n&=&aK_2(\bar{\beta})/\bar{\beta},\\ \label{eq:pmb} 
p&=&amK_2(\bar{\beta})/\bar{\beta}^2=n\,k_BT,\\ \label{eq:r-3nmb}
\rho-3p&=&amK_1(\bar{\beta})/\bar{\beta},
\eq
where $a=4\pi gm^3e^{\beta\mu}/h^3$. Note that the ideal gas equation of state
(\ref{eq:pmb}) is a relativistic invariant.

For the internal energy per particle we then find
\bq
u(T)=\frac{\rho}{n}=m\frac{K_1(\bar{\beta})}{K_2(\bar{\beta})}+3k_B T
=\left\{\begin{array}{ll}\displaystyle
m\left[1+\frac{3}{2}\frac{k_BT}{m}+\ldots\right] & \bar{\beta}\gg 1,\\
3k_BT & \bar{\beta}\ll 1,
\end{array}\right.
\eq
where we used the asymptotic expansions (\ref{eq:asy1}) and (\ref{eq:asy2}).

For the ratio of the specific heats $\gamma(T)=c_p/c_v$ we then find
\bq
\gamma(T)&=&\frac{\left.\frac{du}{dT}\right|_p}{\left.\frac{du}{dT}\right|_v}
=1+\frac{k_B}{\left.\frac{du}{dT}\right|_v}=
\left\{\begin{array}{ll}\displaystyle
5/3 & \bar{\beta}\gg 1,\\
4/3 & \bar{\beta}\ll 1.
\end{array}\right.
\eq

\red{
\subsubsection{Quantum statistics ($\varepsilon=\pm 1$) \cite{Fantoni17d}} 

In our Ref. \cite{Fantoni17d} we did the calculation for the
quantum statistics of identical particles which require either a 
symmetrization (for Bose-Einstein statistics) or antisymmetrization
(for Fermi-Dirac statistics) of the free distinguishable particles
density matrix. Introducing the fugacity $z=e^{\beta\mu}$, for the 
Bose-Einstein statistics we find 
\bq \label{Prb}
\beta p&=&\frac{gm^2}{2\pi^2\beta\hbar^3}\sum_{\nu=1}^\infty
\frac{z^\nu}{\nu^2}K_2(\beta m\nu)~,\\ \label{denrb}
n&=&\frac{gm^2}{2\pi^2\beta\hbar^3}\sum_{\nu=1}^\infty
\frac{z^\nu}{\nu}K_2(\beta m\nu)~,
\eq
and for the Fermi-Dirac statistics 
\bq \label{Prf}
\beta p&=&\frac{gm^2}{2\pi^2\beta\hbar^3}\sum_{\nu=1}^\infty
\frac{(-1)^{\nu-1}z^\nu}{\nu^2}K_2(\beta m\nu)~,\\ \label{denrf}
n&=&\frac{gm^2}{2\pi^2\beta\hbar^3}\sum_{\nu=1}^\infty
\frac{(-1)^{\nu-1}z^\nu}{\nu}K_2(\beta m\nu)~.
\eq
In Ref. \cite{Fantoni17d} we also give the results for the 
non relativistic limit 
$\epsilon(\pp)=\sqrt{\pp^2+m^2}\approx \pp^2/2m$ and the 
extreme relativistic limit $\epsilon(k)\approx|\pp|$ of these expressions.

In the zero temperature limit ($\beta\to\infty$) Eqs. (\ref{Prf}) and 
(\ref{denrf}) reduce to (see \S 2.3 of Ref. \cite{Shapiro-Teukolsky})  
\bq \label{Prdf1}
p&=&\frac{g}{2}\frac{m}{\slashed{\lambda}^3}\phi(x)~,\\ \label{Prdf2}
n&=&\frac{g}{2}\frac{x^3}{3\pi^2\slashed{\lambda}^3}~,\\ \label{Prdf3}
\phi(x)&=&\frac{1}{8\pi^2}\left[x\sqrt{1+x^2}\left(
\frac{2}{3}x^2-1\right)+\ln\left(x+\sqrt{1+x^2}\right)\right]~,
\eq
where $\slashed{\lambda}=\hbar/m$, with $m$ the electron mass, is the
electron Compton wavelength.
}

\section{Statistical mechanics approach}
\label{sec:statm}

The statistical mechanics approach is based on a {\sl many body} distribution 
function \cite{Hansen,Gravitation}.

\subsubsection{Thermal equilibrium of the many bodies}

The aim of equilibrium statistical mechanics is to calculate observable properties 
of a system of interest either as averages over a phase trajectory 
(the method of Boltzmann), or as averages over an ensemble of systems, each 
of which is a replica of the system of interest (the method of Gibbs). In Gibbs 
formulation of statistical mechanics the equilibrium probability distribution for
the systems of $N$ identical bodies of the ensemble is described by $\rho_0^{(N)}$, 
a phase space probability density, in a $6N$ dimensional phase space
$\prod_{i=1}^Nd\xx_i\,d\pp_i=d^NV\,d^N\Pi$, in the classical case or a 
density matrix in the quantum case. 
Here we will only consider the more general quantum case that 
reduces to the classical case at high temperature. 
We will then have \cite{Hansen,Ceperley1995,Becattini2012} 
\red{for distinguishable bodies}
\bq \label{eq:edf}
\rho_0^{(N)}&=&\exp\left(-\int_{\partial\Omega}
T^{\mu\nu}\beta_\nu\,dS_\mu\right),\\ \label{eq:pf}
Z_N&=&{\rm tr}\left(\rho_0^{(N)}\right),
\eq
where $\partial\Omega$ is a general, arbitrary, spacelike hypersurface bounding 
the 4-volume $\Omega$ and $\vect{\beta}(\xx)$ is a 4-vector such that 
$\beta=\sqrt{\beta_\mu\beta^\mu}$ and as usual $1/k_B\beta(\xx)=T(\xx)$ the invariant 
absolute temperature, i.e. the temperature measured by a comoving thermometer.
$Z_N$ is the canonical partition function where ${\rm tr}(\ldots)$ denotes a trace
that requires a path integral in position representation \cite{Ceperley1995}
\red{where for bosons one needs to symmetrize the density
matrix for distinguishable bodies over permutations of their 
$\{\vect{\rr}_i\}$  positions and for fermions one needs to 
antisymmetrize it.} 
\red{So that} $\int T^{00}\,dV=\calh$ with $\calh=\calk+\calv$ the Hamiltonian 
operator of the fluid where $\calk$ is the kinetic energy operator of the $N$ 
bodies \red{and $\calv$ their potential energy}. The covariant form of Eq. 
(\ref{eq:edf}) of the equilibrium statistical 
operator was first used by Weldon \cite{Weldon1982} for the Belinfante 
symmetrized stress energy tensor.

It is then possible to define the $n$-body reduced equilibrium distribution 
functions as
\bq
f_0^{(n)}(\vect{\xx}_1,\ldots,\vect{\xx}_n)&=&\left\langle\left[
\prod_{i=1}^n\sum_{j=1}^N\delta(\vect{\xx}_i-\vect{\rr}_j)
\right]_{\rm DP}\right\rangle\\
&=&\frac{1}{Z_N}\bigintss
\left[\prod_{i=1}^n\sum_{j=1}^N\delta(\vect{\xx}_i-\vect{\rr}_j)
\right]_{\rm DP}
\rho_0^{(N)}(\{\vect{\rr}_k\},\{\vect{\rr}_k\};\beta)
\,d^{4N}\vect{\rr},
\eq
where the thermal average of an operator $\calo$ is 
$\langle\calo\rangle={\rm tr}\left(\rho_0^{(N)}\calo\right)/Z_N$, 
$d^{4N}\vect{\rr}=\prod_{i=1}^Nd\vect{\rr}_i$, 
$\rho_0^{(N)}(\{\vect{\rr}_k\},\{\vect{\rr}_k'\};\beta)$ is the position 
representation of the density matrix (\ref{eq:edf}) at an inverse 
temperature $\beta$ that results from a path integral \cite{Ceperley1995}
\red{with the proper symmetrization or antisymmetrization necessary to
reflect the permutational properties of the identical bodies} 
and the subscript DP means that only the products of Dirac delta 
functions relative to Different Particles should be considered. Here 
$f_0^{(1)}=\int f_0\,d\Pi$ where $f_0$
is the one body distribution function of the previous Section 
\ref{sec:kinetict}.

In a recent project \cite{Fantoni23a} we studied an electron gas at low 
temperatures, the {\sl Jellium}, on the surface of a sphere through the 
path integral Monte Carlo method. A unit sphere is {\sl the} 
surface 
\footnote{Being a manifold of dimension $2<3$ it is conformally flat. 
Moreover in a two dimensional world it is possible to conceive 
anyonic statistics \cite{Lerda} for identical but impenetrable bodies.
For anyons, unlike bosons and fermions the statistics depends on the whole
imaginary time evolution and braiding properties of the path and not just
on its initial and final point. The braid group was introduced in 1925 by 
Emil Artin.} 
of constant positive scalar curvature $2$. In particular we noticed as 
the simulation ``speed'' of the path in a neighborhood of the poles
diminishes.
This is a consequence of the {\sl hairy ball theorem}, according to which 
her Euler class is the obstruction to her tangent planes, the {\sl tangent 
bundle},
\footnote{A particular {\sl fiber bundle}.}
 having always a non vanishing {\sl fiber}, or hair, for any 
{\sl section}
\footnote{In topology, a cross {\sl section} of a fiber (tangent) 
bundle space, $B\times F$ is a graph over 
the {\sl base space} $B$, in this case the sphere. 
A choice of a tangent vector to any point of the sphere 
is a section of the tangent bundle of the sphere.}.
The theorem was first proven by Henri Poincar\'e for 
the sphere in 1885 \cite{Poincare1885}, and extended to higher even 
dimensions in 1912 by Luitzen Egbertus Jan Brouwer \cite{Brouwer1912}.
The theorem has been expressed colloquially as ``you can't comb a 
hairy ball flat without creating a cowlick'' or ``you can't comb 
the hair on a coconut''.
If $z$ is a continuous function that assigns a vector in the three 
dimensional space to every point $\calp$ on a sphere such that $z(\calp)$ 
is always tangent to the sphere at $\calp$, then there is at least one 
pole, a point where the field vanishes, i.e. a $\calp$ such that
$z(\calp)=0$. Every zero of a vector field has a (non-zero) {\sl index}
\footnote{The index of a bilinear function/al is the dimension of the
space on which it is negative definite. According to Morse theorem, 
from the calculus of variations, there is a relation between the
conjugate points (a point of the path where the path cease to be a minimum
of the action) along a classical path to the negative eigenvalues 
of $\delta^2 S$, where $S$ is the action in the path integral. 
More precisely Morse index theorem states that, for
an extremum $\vect{\rr}(t), 0<t<\beta$, 
the index of $\delta^2S$ is equal to the number of conjugate points 
to $\vect{\rr}(0)$ along the path $\vect{\rr}(t)$ (each such conjugate 
point is counted with its multiplicity) \cite{Schulman}. 
In the context of vector fields on a Riemannian manifold
the index is equal to $+1$ around a source or a sink, and more generally equal to 
$(-1)^k$ around a saddle that has $k$ contracting dimensions and $n-k$ 
expanding dimensions.}, 
and it can be shown that the sum of all of the indexes at all of the 
zeros must be two, because the Euler characteristic of the sphere is two. 
Therefore, there must be at least one zero. This is a consequence of the 
{\sl Poincar\'e-Hopf theorem}. The theorem was proven for two dimensions 
by Henri Poincar\'e and later generalized to higher dimensions by Heinz 
Hopf \cite{Hopf1926}. In particular we see how, even a single free particle 
have a path which will be subject to some anisotropy due to the effective 
potential induced by the curvature of the sphere. This effect was studied 
in Refs. \cite{Fantoni23a}.

\subsubsection{Thermal equilibrium of the metric tensor}

A different story is to move the temperature from the stress energy tensor
to the metric tensor as is done in Refs. \cite{Fantoni23b,FantoniCQ} and
in the trilogy \cite{Fantoni24f,Fantoni25a,Fantoni25g} also applied to study 
the vacuum in cosmic space in Ref. \cite{Fantoni25m}. That is, to move the
statistical physics description from the right hand side of Einstein field
equations
\bq \label{eq:efe}
G_{\mu\nu}=\frac{8\pi G}{c^4}T_{\mu\nu},
\eq
to the left hand side. Of course the two descriptions has to give
the same picture. In Ref. \cite{Fantoni24f} we took the statistical 
average of the trace of Einstein field equations
\bq \label{eq:avtefe}
\langle -R\rangle_g=\kappa\langle T_\mu{}^\mu\rangle_t.
\eq
where $\kappa=8\pi G/c^4$ and $R=-G_\mu{}^\mu$ is the scalar curvature,
$\langle\ldots\rangle_g$ is a statistical average on the metric tensor,
and $\langle\ldots\rangle_t$ is a time average. On the right hand side,
replacing the time average with an ensemble average, we find 
\cite{Becattini2012}
\bq
\langle z_\mu T^{\mu\nu}\rangle_t=\frac{
{\rm tr}\left(\rho_0^{(N)}z_\mu T^{\mu\nu}\right)}{Z_N}=
-\frac{\delta}{\delta\beta_\nu(\xx)}\ln Z_N.
\eq
In the above formula, while the left hand side depends on a arbitrary
vector $\vect{z}$, the right hand side is not manifestly dependent on it. 
In fact, the functional derivative of $Z_N$ of Eq. (\ref{eq:pf}) 
includes a hidden dependence on the normal vector as the functional 
derivation implies the choice of a measure, hence of a hypersurface 
and a corresponding normal vector. We will also have
\bq
\langle T_\mu{}^\mu\rangle_t=-\frac{\delta}{\delta\beta}
\ln Z_N.
\eq
Then the virial theorem of Eq. (\ref{eq:avtefe}) can be rewritten as 
\bq
\langle R\rangle_g=\kappa\frac{\delta}{\delta\beta}
\ln Z_N,
\eq
where ${\rm tr}(\ldots)$ denotes a trace, and $\Lambda=\sqrt{4\pi\lambda\beta}$ 
is the de Broglie thermal wavelength, with $\lambda=\hbar^2/2m$. 

For an ideal gas, where the bodies are non interacting, $\calv=0$, we 
immediately find $Z_N=V^N/\Lambda^{-3N}N!$ where 
$\Lambda=\sqrt{4\pi\lambda\beta}$ is the de Broglie thermal wavelength, 
with $\lambda=\hbar^2/2m$. Then 
\bq
\langle R\rangle_g=\kappa\frac{1}{V}\frac{\partial}{\partial\beta}
\ln\left[\frac{1}{N!}\left(\frac{V}{\Lambda^3}\right)^N\right]=
-3n\kappa\frac{\partial}{\partial\beta}\ln\Lambda=
-\frac{3n}{2\beta}\kappa,
\eq
where the functional derivative has been replaced by a partial derivative 
and $V/\Lambda^3$ is the single particle translational partition function, 
familiar from elementary statistical mechanics.

In Ref. \cite{Fantoni24f} we defined a {\sl virial} inverse temperature 
$\tilde{\beta}$ stemming from the thermal fluctuations of the metric 
tensor, as
\footnote{Note that the path integral needed for the calculation
of the left hand side of Eq. (\ref{eq:avtefe}) in the metric tensor 
required the choice of euclidean time whereas the one in the
right hand side requires real time.}
\bq
\tilde{\beta}^{-1}(\xx)=\frac{\tilde{v}}{4}\langle T_\mu{}^\mu\rangle_t,
\eq
where $\tilde{v}$ is a positive constant. Therefore we find the following 
equivalence
\bq
\tilde{\beta}^{-1}(\xx)=\frac{3n\tilde{v}}{8}\beta^{-1}(\xx),
\eq
or 
\bq \label{eq:tep}
\tilde{T}(\xx)=T(\xx),
\eq 
with
\bq
\tilde{k}_B=\frac{3n\tilde{v}}{8}k_B,
\eq
where $n\tilde{v}$ is an intensive quantity, if fact $\tilde{v}$ is a
local volume \cite{Fantoni24f}. In Ref. \cite{Becattini2012}
it was also shown that at thermodynamic equilibrium $\beta^\mu(\xx)$ must be a 
killing vector of the manifold so it must be a constant four vector. Then
$T$ should be independent of $\xx$ at thermodynamic equilibrium.

\subsubsection{Thermal equivalence principle}

This equivalence proves that Einstein field equations offer a symmetric
way to study statistical physics where one can either work at the level
of the many body system encoded in the stress energy tensor on the right hand 
side of (\ref{eq:efe}) or at the level of the thermal fluctuations of 
the metric tensor on the left hand side of (\ref{eq:efe}). The two 
descriptions are {\sl equivalent}. This is a {\sl thermal equivalence 
principle} in physics: ``given a many body system in general relativity
its thermal equilibrium properties derived from its statistical physics 
description are equivalent to the properties of a statistical physics
description of the metric of the spacetime that it influences and
viceversa''. In other words: ``A statistical ensemble of bodies goes into 
thermal equilibrium with the spacetime it occupies''. This is established
by Eq. (\ref{eq:tep}). 

\red{It is nice to see how the statistical physics theory and the general 
relativity theory can be made to be consistent one with the other. Even if 
this fact can be considered superfluous in my opinion it cannot be given 
for granted or even treated carelessly and it is not at all trivial. 
This thermal equivalence principle issues a bridge between the two 
communities of statistical physics and general relativity physics.}

\red{
This principle should be taken into account for example when constructing
of a more accurate and realistic equation of state of a White Dwarf which 
requires dropping the assumptions of an ideal and perfect electron gas. 
This has been accomplished with various mean field theories well
illustrated in section \S 2 of the book of Shapiro and Teukolsky
\cite{Shapiro-Teukolsky}. Since for a White Dwarf the Wigner-Seitz radius 
is samll, $0.0003<r_s<0.01$
\footnote{The lower bound is dictated by the requirement of being below 
neutron drip. The upper bound can be inferred 
from figure 3.2 of the book \cite{Shapiro-Teukolsky} taking a star mass as
low as $0.7 M_\odot$.}, the corrections due to the Coulomb interaction 
will be small. 
The principal effect of the electrostatic corrections is to give smaller radii 
and larger central densities compared with Chandrasekhar's models of the 
same mass.
Nonetheless, in order to make further progress towards an  
even more accurate equation of state many body methods are necessary. This
is not simply an academic exercise because the stars are probably the
objects more observed and measured in nature and we can hope to better
understand the laws of nature by comparing our theories on earth with 
the data from astronomical observations. Until recently stars were
only present in the sky. Only recently we are able to create a star
artificially in a earthly laboratory \cite{wikifusion}. Moreover we may hope 
to be able to detect some gravitational wave generated by binary systems 
of White Dwarfs as they spiral closer to an eventual merger. These systems, 
which can be detached or interact through mass transfer, are major sources 
of gravitational waves for future detectors like the 
Laser Interferometer Space Antenna (LISA) planned by the European Space Agency
(ESA) \cite{wikilisa}. One may then start for example from the properties of Jellium 
\cite{Kenny1996,Brown2013,Fantoni21b} where the ions component 
is approximated by a uniform neutralizing background. This is just a first
brute approximation to the more realistic model of a two component plasma
\cite{Ceperley1981,Ceperley1987}
\footnote{Note that since the mass of a proton is about $1000$ electron 
masses the ions component diffusion would 
be $1000$ times slower making it much more classical in a first 
principles statistical physics description.}, 
but even so it poses the extremely challenging problem of the determination 
of the ground state or non zero temperature properties of a many electron 
system on a curved 
spacetime \cite{Fantoni18c,Fantoni23a} with the additional subtleties of 
overcoming the fermion sign problem \cite{Ceperley1980,Ceperley1991,Ceperley1995} 
and ordering problems on properly self adjoint operators subject to holonomic 
constraints as the ones necessary in a quantum theory of curved spacetime
\cite{Fantoni23b,Fantoni24f,Fantoni25a,Fantoni25g}.
The effects of temperature and/or general relativity on the global structure of 
a White Dwarf have been recently studied in greater depth considering realistic 
models of dense matter \cite{Boshkayev2016,Carvalho2018,Nunes2021}. 
Moreover, full recent evolution simulations of the most massive White Dwarfs 
have been also carried out \cite{Althaus2022,Althaus2023}. Our thermal 
equivalence principle could offer additional insights through the 
equivalent treatment of the statistical physics description of the 
metric of spacetime. 
}
\section{Conclusions}
\label{sec:conclusions}

In this work we determined a thermal equivalence principle for a 
statistical theory of gravitation. First of all it is important to realize 
that at low temperatures a statistical theory of gravity will necessarily
put together the quantum world with our Universe ruled by general relativity.
\red{This is certainly relevant for a refinement of the existing equations
of state of stellar interiors which allow the prediction and better 
understanding of stellar evolution}. Moreover, outer space, or simply space, 
is the expanse that exists beyond Earth 
atmosphere and between celestial bodies. It contains very low particle 
densities, constituting a near perfect vacuum of predominantly hydrogen and 
helium plasma, permeated by electromagnetic radiation, cosmic rays, 
neutrinos, magnetic fields and dust. The baseline temperature of outer space, 
as set by the background radiation from the Big Bang, is $\approx 2.7$ K. 
Intergalactic space takes up most of the volume of the universe, but even 
galaxies and star systems consist almost entirely of empty space. Most of the 
remaining mass-energy in the observable universe is made up of an unknown 
form, dubbed dark matter (60\% of the Universe) and dark energy 
(27\% of the Universe).

The program of constructing a well defined statistical theory of our
Universe is one of the greatest challenges of contemporary physics 
which had been foreseen by Einstein in his renown iconic phrase ``God doesn't
play dice''. From the point of view of the challenge that it
offers to mathematics one needs a way to create a bridge between the 
variational theory of functional integrals or more specifically path integrals 
and differential geometry or more specifically Riemannian geometry. 
From this point of view it seems natural to predict that differential 
topology will play a crucial role. Recently we carried out some path integral
(Monte Carlo) simulations for Jellium (an electron plasma at low temperature)
on the surface of a sphere, probably the simplest of all curved smooth 
manifolds. And already in that study we found important topological effects
on the electrons paths. It is important to realize that this kind of 
calculations can be considered as toy simulations for a many body system
on a more complex smooth manifold as needed by spacetime in general relativity.

In this work we show that the temperature as defined in this kind of 
statistical physics studies of many bodies plays the same role as the one that 
can be defined by a path integral on the spacetime metric that was introduced 
in Ref. \cite{Fantoni24f}
\footnote{
\red{
Some may criticize our Ref. \cite{Fantoni24f} in that the temperature 
should be considered as a property ``external'' to spacetime. But our 
temperature measures the thermal fluctuations of the metric tensor itself 
as illustrated in Ref. \cite{Fantoni25a} where the foundations of our 
statistical physics formulation of gravity were demonstrated.}
}. 
This is perfectly natural from the point of view of 
the linear constraint given by the Einstein field equations. This symmetry 
between the statistical physics of many body matter in the Universe and a 
statistical physics theory of the metric tensor where matter lives, 
offers naturally a thermal equivalence principle stating that the 
material and the spacetime are in thermal equilibrium one
another. \red{One could either let the classical many bodies live in a 
quantum spacetime or a quantum many bodies live in a classical spacetime,
i.e. move the path integral description from left to right in Einstein
field equations. The two approaches have to coincide.}


\section*{Author declarations}

\subsection*{Conflicts of interest}
None declared.

\subsection*{Data availability}
The data that support the findings of this study are available from the 
corresponding author upon reasonable request.

\subsection*{Funding}
None declared.

\bibliography{mbgr}

\end{document}